\documentstyle[epsfig]{mn}
\begin{document}
\input BoxedEPS.tex
\SetEPSFDirectory{./}
\SetRokickiEPSFSpecial
\HideDisplacementBoxes

\title
[ELAIS Paper IV: the preliminary $90\mu$m luminosity function
]
{The European Large Area ISO Survey IV: 
the preliminary $90\mu$m luminosity
function}

\author[Stephen Serjeant, et al]{
S. Serjeant$^{1}$,  
A. Efstathiou$^{1}$, 
S. Oliver$^{1,2}$, 
C. Surace$^{1}$, 
P. Heraudeau$^{3}$, 
  \vspace*{0.2cm}\\ {\LARGE
M. Linden-V\o rnle$^{4}$, 
C. Gruppioni$^{5}$,
F. La Franca$^{6}$, 
D. Rigopoulou$^{7}$, 
  }\vspace*{0.2cm}\\ {\LARGE
T. Morel$^{1}$, 
H. Crockett$^{1}$, 
T. Sumner$^{1}$, 
M. Rowan-Robinson$^{1}$,
M. Graham$^{1}$
 }\vspace*{0.2cm}\\
$^{1}$Astrophysics Group, Blackett Laboratory, Imperial College, 
Prince Consort Road,London SW7 2BW, UK\\
$^2$ Astronomy Centre, CPES, University of Sussex,
Falmer, Brighton BN1 9QJ\\
$^3$ Max-Planck-Institut f\"{u}r Astronomie, K\"{o}nigstuhl 17, D69117, 
 Heidelberg, Germany\\
$^4$ Danish Space Research Institute, 30 Juilane Maries Vej, DK-2100
Copenhagen 0, Denmark\\
$^5$ Osservatorio Astronomico di Bologna, via Ranzani 1,
40127 Bologna, ITALY\\
$^6$ Dipartimento di Fisica E. Amaldi, Via della Visca Navale 84,
 Rome, Italy\\
$^{7}$ Max-Planck-Institut f\"{u}r extraterrestrische Physik
Giessenbachstra$\beta$e, 85748 Garching, Germany\\
}
\date{Accepted;
      Received;
      in original form 2000 Jun 19}
 
\pagerange{\pageref{firstpage}--\pageref{lastpage}}
\pubyear{2000}
\volume{}

\label{firstpage}

\maketitle


\begin{abstract}
We present the luminosity function of $90\mu$m selected galaxies from
the European Large Area ISO Survey (ELAIS), extending to $z=0.3$.
Their luminosities 
are in the range 
$10^{9}<h_{65}^{-2}L/L_\odot<10^{12}$, i.e. non-ultraluminous. 
From our sample of $37$ reliably detected galaxies in the ELAIS S1
region from the Efstathiou et al. (2000) $S_{90}\ge 100$mJy database,
we found optical, $15\mu$m or $1.4$GHz 
identifications for $24$ ($65\%$). We have obtained
2dF and UK Schmidt FLAIR
spectroscopy of $89\%$ of IDs to rigid multivariate flux limits. 
We construct a luminosity function
assuming (a) our spectroscopic subset is an unbiased
sparse sample, and
(b) there are no galaxies which would not be represented in our
spectroscopic sample at {\it any} redshift. We argue that we can be
confident of both assumptions. 
We find the luminosity function is well-described by the local
$100\mu$m luminosity function of Rowan-Robinson, Helou \& Walker
(1987). 
{\it Assuming}
this local normalisation, we derive luminosity
evolution of $(1+z)^{2.45\pm0.85}$ ($95\%$ confidence). 
We argue that star formation dominates the bolometric luminosities of these 
galaxies and 
we derive comoving star formation rates in broad agreement with the
Flores et al. (1999) and Rowan-Robinson et al. (1997) mid-IR-based
estimates.  

\end{abstract}

\maketitle

\section{Introduction}\label{sec:introduction}
The study of the star formation history of the Universe is an extremely
active field, in which much of the current debate is centred on the
uncertainties in dust obscuration. Selecting star forming galaxies in
the UV is relatively cheap in observing time, but 
such selection makes the samples extremely sensitive to dust
obscuration. Detecting the reprocessed starlight 
requires sub-mm surveys (e.g. Smail, Ivison \& Blain 1997; 
Hughes et al. 1998; Barger et al. 1998, 1999; Eales et al. 1999, 
Peacock et al. 2000;
Ivison et al. 2000a,b, Dunne et al. 2000) and/or 
space-based mid-far-IR surveys (e.g. 
Rowan-Robinson et al. 1997,
Taniguchi et al. 1997, Kawara et al. 1998, Flores et al. 1999, 
Puget et al. 1999, Oliver et al. 2000; see e.g. Oliver 2000 for a
review). 
As such, the European Large Area ISO Survey (ELAIS) 
is well-placed for the study of 
the evolution and obscuration of the star formation in the Universe, 
and strong constraints at $z\stackrel{<}{_\sim}1$ 
are possible from ELAIS (Oliver et al. 2000). 

\begin{table*}
\begin{tabular}{llllllllll}
Name  & RA          & Dec          & $S_{90}$/Jy & $S_{15}$/mJy & $S_{1.4}$/mJy & R & z & z$_{\rm max}$ & $L/L_\odot$ \\
      & (J2000)     & (J2000)      &             &          &           &   &   &             \\
ELAISP90\_J003857-424358        & 00 38 57.7 & -42 43 58.8  &   0.43 &     2.9 &    0.964    &             &          & &         \\
ELAISP90\_J003056-441633        & 00 30 57.0 & -44 16 33.6  &   0.35 &     4.9 &             &     13.7    &   0.019  &0.0351$^P$ & $2.8\times 10^{9}$\\ 
ELAISP90\_J003510-435906        & 00 35 10.5 & -43 59  6.0  &   0.41 &     9.6 &             &     12.1    &   0.136  &0.272$^P$ & $1.7\times 10^{11}$\\ 
ELAISP90\_J003431-432613        & 00 34 31.6 & -43 26 13.2  &   0.59 &         &             &     14.0    &   0.053  &0.127$^P$ & $3.7\times 10^{10}$\\ 
ELAISP90\_J003914-430437        & 00 39 14.8 & -43 04 37.2  &   1.79 &    16.2 &             &     11.6    &   0.012  &0.049$^P$ & $5.7\times 10^{9}$\\ 
ELAISP90\_J003021-423657        & 00 30 21.1 & -42 36 57.6  &   1.17 &    22.3 &             &     16.0    &   0.149  &0.493$^P$ & $6.0\times 10^{11}$\\ 
ELAISP90\_J003459-425718        & 00 34 59.0 & -42 57 18.0  &   0.55 &     5.5 &     1.371   &     14.9    &   0.055  &0.101$^R$ & $3.7\times 10^{10}$\\ 
ELAISP90\_J003134-424420        & 00 31 34.5 & -42 44 20.4  &   0.31 &         &             &     13.1    &   0.027  &0.048$^P$ & $5.1\times 10^{9}$\\ 
ELAISP90\_J003242-423314        & 00 32 42.9 & -42 33 14.4  &   0.24 &     2.6 &    0.463    &     15.2    &   0.053  &0.081$^P$ & $1.5\times 10^{10}$\\ 
ELAISP90\_J003615-424344        & 00 36 15.4 & -42 43 44.4  &   0.31 &         &             &     17.2    &   0.115  & &        \\ 
ELAISP90\_J003516-440448        & 00 35 16.1 & -44 04 48.0  &   0.29 &         &             &             &          & &        \\
ELAISP90\_J003741-440227        & 00 37 41.6 & -44 02 27.6  &   0.21 &         &             &     19.9    &   0.348  & &        \\ 
ELAISP90\_J003431-433806        & 00 34 31.2 & -43 38  6.0  &   0.75 &         &    0.595    &     18.4    &   0.199  &0.357$^O$ & $6.8\times 10^{11}$\\ 
ELAISP90\_J003405-423816        & 00 34  5.3 & -42 38 16.8  &   0.23 &         &             &             &          & &        \\
ELAISP90\_J003358-441102        & 00 33 58.2 & -44 11  2.4  &   0.23 &     5.0 &             &     18.9    &   0.305  &0.446$^O$ & $5.2\times 10^{11}$\\ 
ELAISP90\_J003501-423852        & 00 35  1.9 & -42 38 52.8  &   0.28 &         &    0.606    &     15.9    &   0.054  &0.090$^P$ & $1.8\times 10^{10}$\\ 
ELAISP90\_J003254-424608        & 00 32 54.2 & -42 46  8.4  &   0.46 &     2.5 &    0.469    &     17.7    &   0.190  &0.207$^C$ & $3.8\times 10^{11}$\\ 
ELAISP90\_J003531-423314        & 00 35 31.4 & -42 33 14.4  &   0.16 &         &             &             &          & &        \\
ELAISP90\_J003635-430148        & 00 36 35.7 & -43 01 48.0  &   0.12 &     4.5 &    0.358    &     21.5    &          & &        \\
ELAISP90\_J003019-432006        & 00 30 19.8 & -43 20  6.0  &   0.20 &         &             &             &          & &        \\
ELAISP90\_J003839-422324        & 00 38 40.0 & -42 23 24.0  &   0.14 &         &             &             &          & &        \\
ELAISP90\_J003014-440509        & 00 30 14.3 & -44 05  9.6  &   0.10 &     3.6 &             &             &          & &        \\
ELAISP90\_J003912-431203        & 00 39 12.2 & -43 12  3.6  &   0.15 &     5.0 &             &     14.8    &          & &        \\
ELAISP90\_J003133-425100        & 00 31 33.9 & -42 51  0.0  &   0.11 &     6.3 &             &     18.0    &   0.211  &0.222$^P$ & $1.1\times 10^{11}$\\ 
ELAISP90\_J003304-425212        & 00 33  4.3 & -42 52 12.0  &   0.19 &     1.4 &             &     15.6    &          & &        \\
ELAISP90\_J003046-432204        & 00 30 46.4 & -43 22  4.8  &   0.11 &     6.2 &             &     15.6    &   0.073  &0.075$^P$ & $1.3\times 10^{10}$\\ 
ELAISP90\_J003027-433050        & 00 30 27.7 & -43 30 50.4  &   0.11 &     4.9 &             &     15.7    &   0.072  &0.074$^P$ & $1.2\times 10^{10}$\\ 
ELAISP90\_J002845-424420        & 00 28 45.5 & -42 44 20.4  &   0.11 &         &             &             &          & &        \\
ELAISP90\_J003719-421955        & 00 37 19.4 & -42 19 55.2  &   0.16 &         &             &             &          & &        \\
ELAISP90\_J003725-424554        & 00 37 25.2 & -42 45 54.0  &   0.12 &         &             &             &          & &        \\
ELAISP90\_J003731-440758        & 00 37 31.3 & -44 07 58.8  &   0.32 &         &             &     20.0    &          & &        \\
ELAISP90\_J003625-441127        & 00 36 25.0 & -44 11 27.6  &   0.14 &         &     4.265   &             &          & &        \\
ELAISP90\_J003721-434228        & 00 37 21.5 & -43 42 28.8  &   0.12 &    17.3 &    0.451    &     17.4    &   0.225  &0.248$^P$ & $1.4\times 10^{11}$\\ 
ELAISP90\_J003348-433032        & 00 33 48.3 & -43 30 32.4  &   0.15 &         &             &             &          & &        \\
ELAISP90\_J003149-423628        & 00 31 49.2 & -42 36 28.8  &   0.11 &         &             &             &          & &        \\
ELAISP90\_J003244-424803        & 00 32 44.4 & -42 48  3.6  &   0.12 &     1.2 &             &     19.0    &  0.192   & &        \\ 
ELAISP90\_J003323-432634        & 00 33 23.2 & -43 26 34.8  &   0.19 &         &             &     18.7    &  0.316   & &        \\ 
\end{tabular}
\caption{\label{tab:sample} ELAIS S1 $90\mu$m sample. The $90\mu$m photometry
has a $\sim30\%$ possible systematic uncertainty (see Paper III), and
the $15\mu$m photometry is subject to a possible systematic scaling to 
fainter fluxes by up to a factor of $\sim1.5$ (see Paper II). We
assume $H_0=65$ km s$^{-1}$ Mpc$^{-1}$, $\Omega_0=1$, $\Lambda=0$, and 
bolometric luminosities are calculated at $90\mu$m assuming a constant
$\nu L_\nu$ spectrum. The maximum accessible redshifts $z_{\rm max}$
quoted assume no evolution, and are labelled respectively as $P$, $R$, 
$O$ and $C$ for maximum redshifts imposed by the $90\mu$m PHOT flux
limit, the radio flux limit, the optical magnitude limit and the CAM
flux limit. 
Two galaxies (ELAISP90\_J003721-434228 and 
ELAISP90\_J003021-423657) are Seyfert II galaxies and two 
further galaxies (ELAISP90\_J003431-433806 and 
ELAISP90\_J003133-425100) have an early type spectrum; all
the remainder show starburst emission line spectra.
Also listed are $4$ redshifts for candidate optical
IDs with the optical magnitudes listed, but which lack other
multi-wavelength detections and are too faint optically to be certain
that the correct ID has been found. All $4$ such identifications show
starburst emission line spectra, but nevertheless
are excluded from the luminosity
function analysis as they fail the strict
selection function defined in the text.
}\end{table*}

The scientific aims and strategy of ELAIS are presented in detail in
Paper I (Oliver et al. 2000). 
In Papers II and III (Serjeant et al. 2000, 
Efstathiou et al. 2000) we presented respectively the ELAIS
Preliminary Analysis source
counts from the CAM ($6.7\mu$m and $15\mu$m) and PHOT ($90\mu$m)
instruments on ISO. 
The $90\mu$m sample covered $11.6$ square degrees, and the source
counts were found to agree well at the bright end with the IRAS
$100\mu$m counts. Excellent agreement was also achieved with a parallel 
independent pipeline (Surace et al. in prep.). 

In this paper we present the first $90\mu$m luminosity function from
our initial spectroscopic campaigns in the ELAIS S1 field 
(Oliver et al. 2000 in prep., Gruppioni et al. 2000 in prep., La
Franca et al. 2000 in prep., Linden-V\o rnle et al. 2000 in prep.). 
Section \ref{sec:method} defines our sample. 
In section \ref{sec:results} we use this
catalogue to derive a $90\mu$m luminosity function. We discuss the
implications of our results in section \ref{sec:discussion}. 
We assume $H_0=65$ km s$^{-1}$ Mpc$^{-1}$, $\Omega_0=1$, $\Lambda=0$
throughout. 




\section{Sample selection and data acquisition}\label{sec:method}
The parent sample for this study is the Preliminary Analysis catalogue
of Efstathiou et 
al. 2000 (Paper III), with $90\mu$m fluxes satisfying
$S_{90}>100$mJy. The completeness of this sample falls approaching
this limit but has been well-quantified with simulations (figure
6 of Paper III). 
We will use the calibration adopted in Paper III. 
The $90\mu$m flux calibration is still uncertain to within $\sim30\%$,
as discussed in Paper III; however, in order to be consistent with the 
$100\mu$m IRAS source counts (assuming $S_{90}\simeq S_{100}$) the flux
calibration has to be very close to 
the Paper III value. Therefore when we compare our $90\mu$m luminosity 
function to the local $100\mu$m luminosity function below, we will
treat the flux calibration as being accurately known. Nevertheless, in 
predicting the redshift distributions of other $90\mu$m surveys, we
will not neglect the flux calibration systematics. 



\begin{figure}
\centering
  \ForceWidth{4in}
 \vspace*{-1cm}
  \hSlide{-1cm}
  \BoxedEPSF{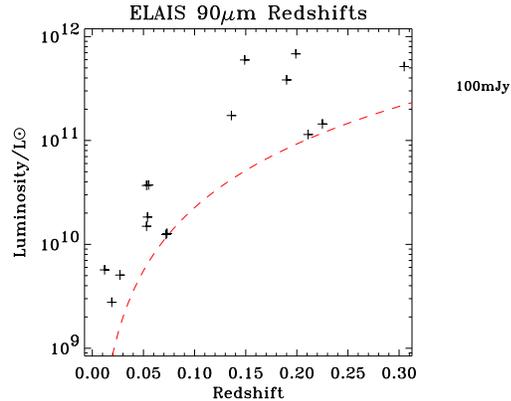}
 \vspace*{-1cm}
\caption{\label{fig:lz}
Luminosity - redshift plane. The $100$mJy flux limit is marked.
}
\vspace*{-0.1cm}
\end{figure}

We restrict our study to the ELAIS S1 field with $90\mu$m, $1.4$GHz
and $15\mu$m survey coverage,
which covers $3.96$ square degrees and has the most extensive available
optical spectroscopy. Large-scale-structure variations are negligible
in the comoving volume of the total ELAIS survey, but the ELAIS S1
field has slightly lower source 
counts than the global average. Consequently we reduced the effective
area by $20\%$ (a factor of $0.8$) to make the S1 counts normalisation
match the whole survey, to account for any possible large scale structure
fluctuations. (This correction is smaller than the errors in the
luminosity function derived below.) 
We sought identifications of this sample using (a) the APM (Automatic
Plate Measuring machine), (b)
sub-mJy decimetric radio sources from our Australia Telescope
Compact Array (ATCA) imaging (Gruppioni et
al. 1999), (c) our ELAIS ISOCAM $6.7\mu$m and $15\mu$m catalogues
(Serjeant et al. 2000, Paper II). By randomising our $37$ catalogue
positions in the cross-correlation, we found that $<1$ optical APM
identifications brighter  
than $R=17$ should occur by chance within the error ellipses of all our
PHOT sources, and similarly restricted the association radius for the CAM
and radio catalogues. Optical spectroscopy was obtained at the 
$2$ degree field (2dF) at
the Anglo-Australian Telescope (AAT),
the Fibre Linked Array Image Reformatter (FLAIR)  
spectrograph at the UK Schmidt and with the 
Danish Faint Object Spectrograph and Camera (DFOSC) at the Danish
$1.54$m at ESO. The optical spectroscopic 
catalogue will appear in future papers in this series. 
Of the $37$ reliably detected galaxies, $24$ had multi-wavelength
identifications of
which $18$ are above the combined multivariate flux limits discussed
above. Of
these we have obtained optical spectroscopy of $89\%$. The ELAIS S1
$90\mu$m sample is presented in table \ref{tab:sample} with the
cross-identifications and available redshifts. This table also lists
redshifts for four identifications for which the identification is
uncertain. These galaxies fail the strict 
selection function defined
below and are excluded from the luminosity function analysis below.
Two galaxies (ELAISP90\_J003721-434228 and
ELAISP90\_J003021-423657) are Seyfert II galaxies and two  
further galaxies  (ELAISP90\_J003431-433806 and 
ELAISP90\_J003133-425100) have an early type spectrum; all 
the remainder (including the four excluded identifications)
show starburst emission line spectra.

\begin{figure*}
\centering
 \vspace*{-1cm}
  \ForceWidth{5in}
  \hSlide{-3.5cm}
  \BoxedEPSF{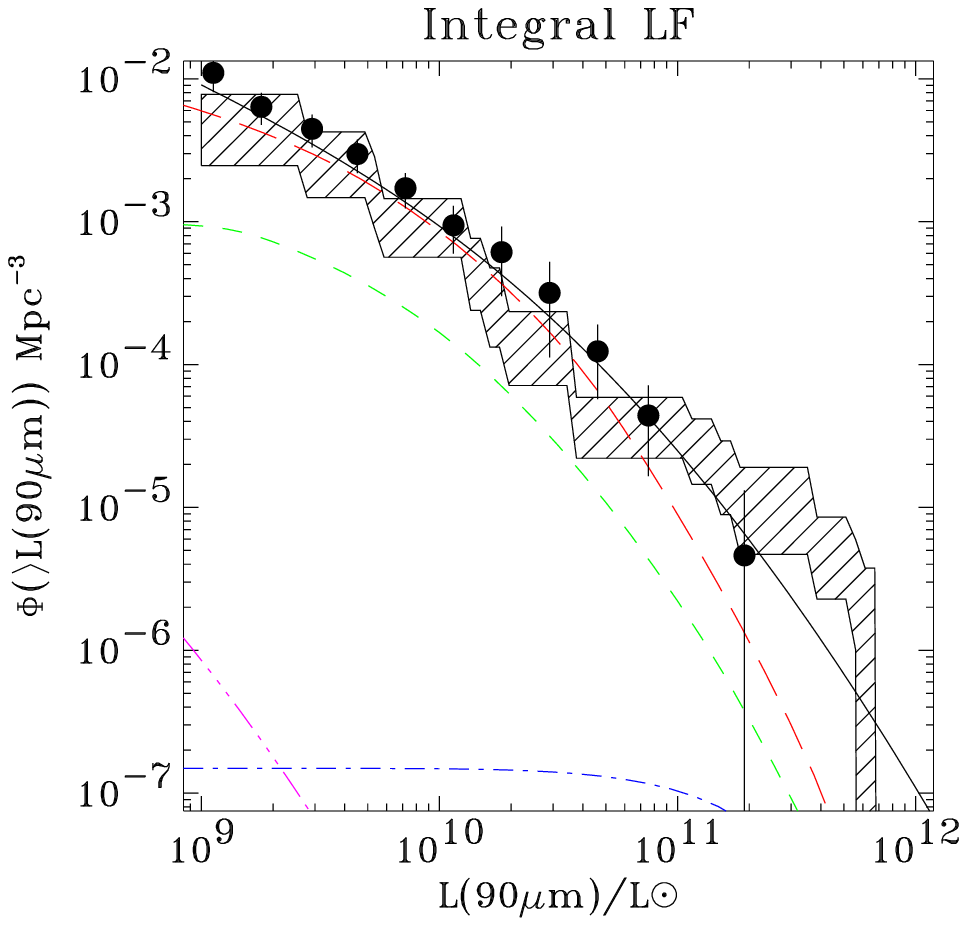}
\vspace*{-9.1cm}
  \ForceWidth{5in}
  \hSlide{4cm}
  \BoxedEPSF{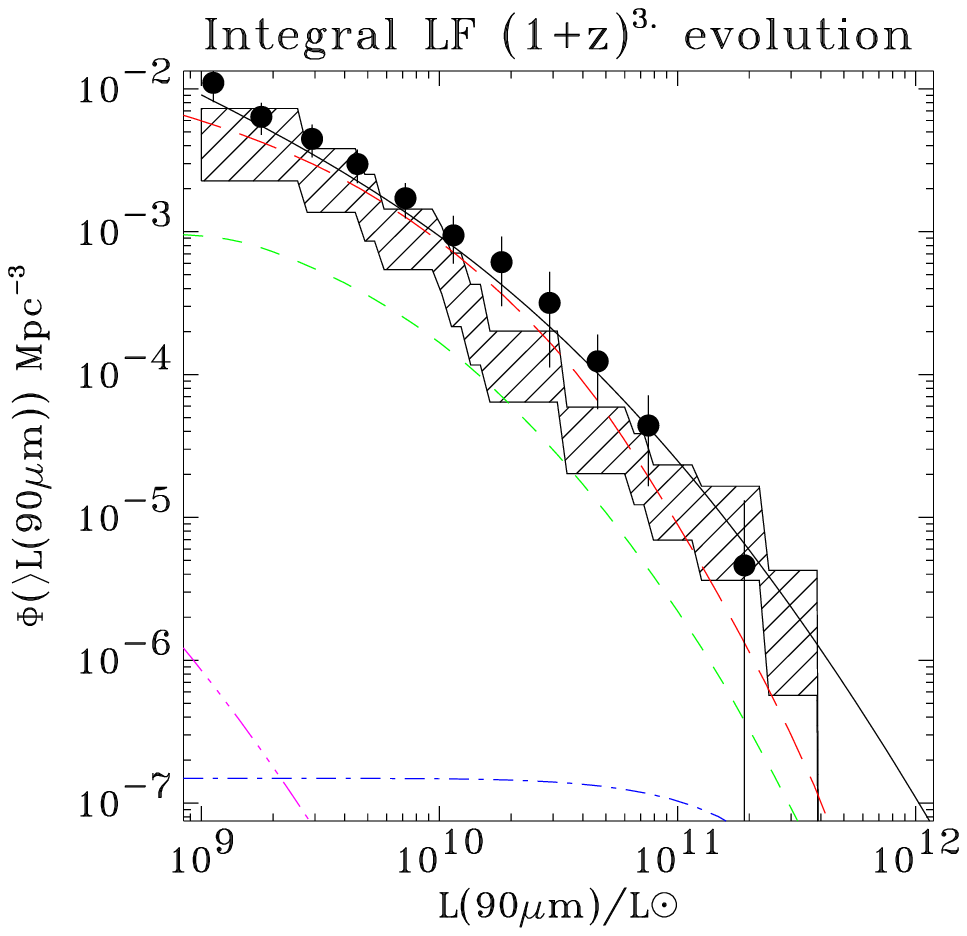}
 \vspace*{-1cm}
 \vspace*{-1cm}
  \ForceWidth{5in}
  \hSlide{-3.5cm}
  \BoxedEPSF{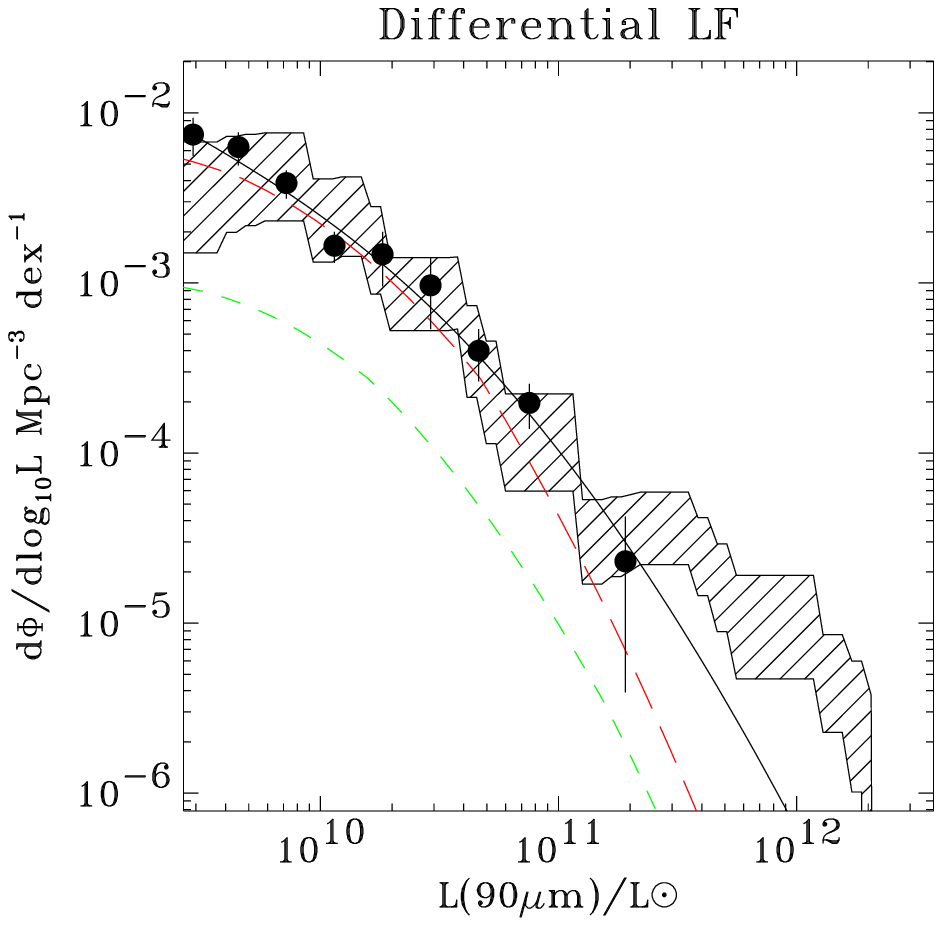}
\vspace*{-9.1cm}
  \ForceWidth{5in}
  \hSlide{4cm}
  \BoxedEPSF{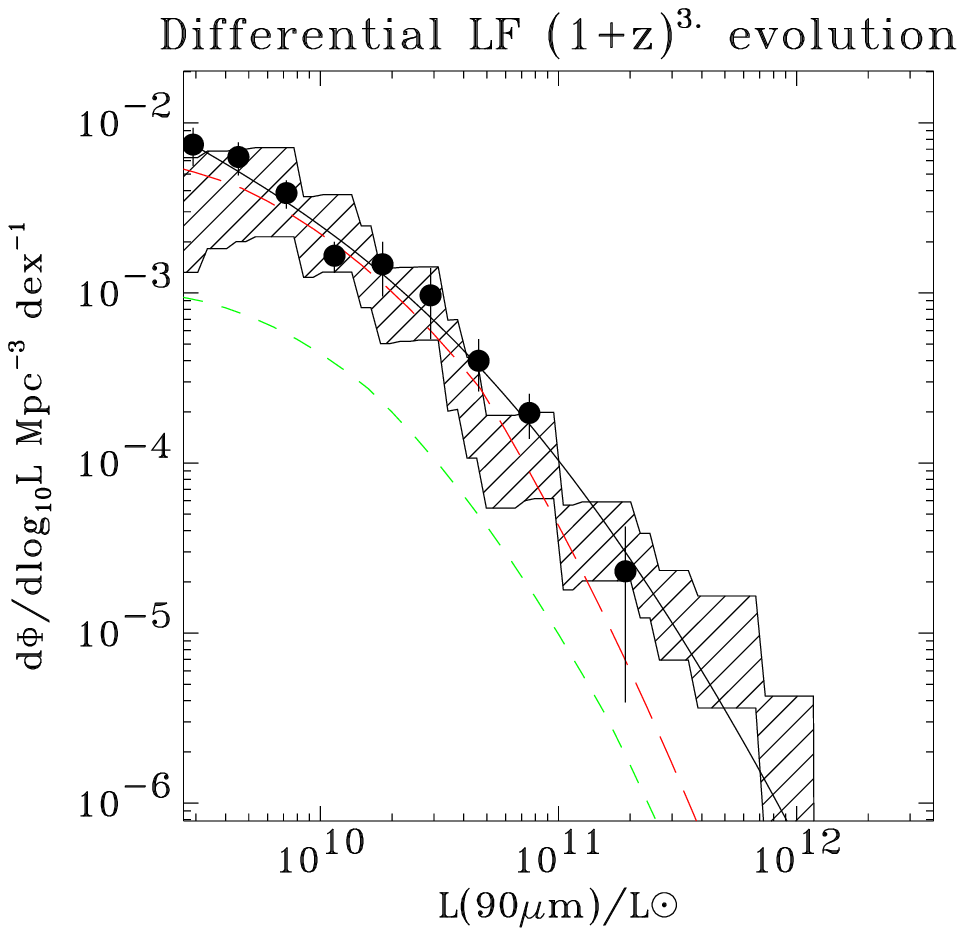}
 \vspace*{-1cm}
\caption{\label{fig:lf}
Luminosity function assuming no evolution (left) and $(1+z)^3$
luminosity evolution (right). Upper figures show the integral
luminosity function, lower figures show the differential form.
Shaded area is the $\pm 1\sigma$ error
region from this study. For the differential counts, the shaded area
indicates the error on the luminosity function in a $\pm 0.5$ dex
bin; the differential counts are therefore effectively smoothed by a
$1$ dex boxcar. 
The data points show the local $100\mu$m
luminosity function of Rowan--Robinson, Helou \& Walker (1987), and
the full line is a fit to this local data. The broken lines show the
populations in the Rowan--Robinson (2000) model. The long dashed line
is the ``cirrus-like'' population, i.e. galaxies with bolometric
luminosities dominated by new stars heating
previously-created dust. N.B., this population is not necessarily
identified with NGC 6090-like cirrus galaxies.
The short dashed line is the starburst
population; dash-dot indicates Arp-220-like galaxies; dash-dot-dot-dot 
shows the AGN population. 
}
\end{figure*}


\section{Results: the $90$ micron luminosity
function}\label{sec:results} 
Our selection function is as follows. In order to be in the spectroscopic
target list, a $90\mu$m galaxy must either have (a) a reliable optical
identification brighter than $R=17$, or (b) a reliable $15\mu$m or
$1.4$GHz identification, together with an optical ID brighter than $R=20$.
The $15\mu$m identifications were made to $S_{15}\ge 2$mJy, and the
radio identifications to $0.2-0.4$mJy depending on position (Gruppioni 
et al. 1999). Only $90\mu$m
galaxies in the area with radio and $15\mu$m survey coverage were
considered. We use the M82 starburst model from 
Rowan-Robinson et al. 1997 to estimate the K-corrections, with
$S_{60\mu\rm m}=120 S_{1.4\rm GHz}$ and a radio spectral index of
$d\log_{10}S/d\log_{10}\nu=-0.8$. 
For optical K-corrections we assume an optical spectral index of $-3$. 
The luminosity-redshift plane is shown in figure 
\ref{fig:lz}, with bolometric luminosities calculated at $90\mu$m 
assuming constant $\nu L_\nu$ (for ease of conversion to other
luminosity scales). Our method for calculating the luminosity function 
in the face of this complicated selection function is dealt with in
the appendix. 
Note that our method relies on the underlying assumption that no
galaxies are missing at {\it all} redshifts 
due to the multivariate flux limits. 
However, we can be
confident of this in our case, as any sufficiently low-redshift galaxy
will have an optical identification that will pass criterion (a) of our
selection function. A hypothetical local population of very optically
faint but far-IR-bright galaxies can already be excluded from IRAS
samples.
We must also assume that our $89\%$ complete optical spectroscopy is a
random sparse 
sample of the total identifications (though not necessarily of the
total $90\mu$m sample). Although with our current
spectroscopic sample this is an a posteriori
selection rather than a priori, there were no selection biases in the
spectroscopic data acquisition which could skew the sample
selection function. We therefore correct our effective areal coverage
by a factor of $0.89$ to account for the small optical spectroscopic
incompleteness. 



The integral luminosity function is given by
\begin{equation}
\Phi(>L) = \sum_{L_i > L} V^{-1}_{{\rm max}, i}
\end{equation}
where the sum is performed over all objects having luminosities
greater than $L$. In figure \ref{fig:lf} 
we show this integral  
luminosity function, and 
compare it to the local $100\mu$m luminosity function derived by
Rowan-Robinson, Helou \& Walker 1987 (binned data). The latter was
derived from a sample in the north Galactic pole, and is subject to
possible large scale structure variations. To estimate this, we
compared the normalisation of their $60\mu$m luminosity function with
that of Saunders et al. (1990), and derived a correction factor of
$0.77$ to the normalisation of their $100\mu$m luminosity function. 
The best-fit function to this local data is plotted as a
full line. The adopted functional form is identical to Saunders et
al. (1990): 
\begin{equation}
\phi(L) = \phi_* (\frac{L}{L_*})^{1-\alpha} 
	\exp(-\frac{1}{2\sigma^2} \log^2(1+\frac{L}{L_*}))
\end{equation}
and the best-fit parameters are given in table \ref{tab:local_lf}. 
Also plotted in this figure are the models of Rowan-Robinson (2000). 


The shaded area shows the 
$\pm1\sigma$ errors from our sample. There is a marginally significant
excess at the highest luminosities. The $\langle V/V_{\rm max}\rangle$ 
statistic for this sample is $0.54\pm0.07$; a Kolmogorov-Smirnov test
on the $V/V_{\rm max}$ distribution shows only a $34\%$ probability
of inconsistency with the top-hat distribution $U[0,1]$. 
Assuming $(1+z)^3$ luminosity evolution (figure \ref{fig:lf})
gives $\langle V/V_{\rm max}\rangle=0.51\pm0.07$, and 
in both the evolving and non-evolving models, a
Kolmogorov-Smirnov test on the observed luminosity distribution
also gives acceptable confidence levels. 
Our
spectroscopic sub-sample on its own is 
therefore not sufficiently large to reliably detect evolution. 



\begin{figure}
\centering
  \ForceWidth{6in}
\vspace*{-3.2cm}
  \hSlide{-3.5cm}
  \BoxedEPSF{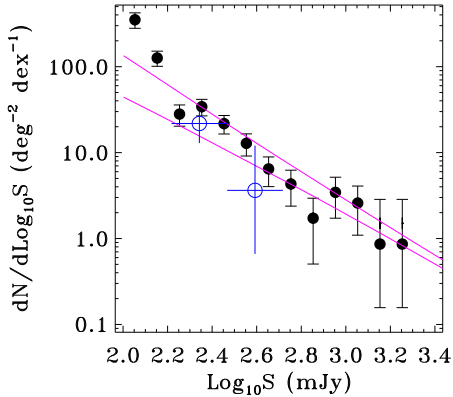}
  \vspace*{-2cm}
\caption{\label{fig:counts}
ELAIS source counts at $90\mu$m, compared to a no-evolution 
model (shallow curve) and the $(1+z)^{3}$ luminosity evolution model
(steeper curve). 
The no-evolution curve has a slightly shallower slope than its
Euclidean equivalent. 
Also shown as open symbols 
are the Lockman Hole counts from Linden-V\o rnle et al. (2000). 
All errors are Poissonian, except in the case of a single object in a
bin for which the $\pm1\sigma$ bounds on the number of objects in the
bin are $0.18-3.3$. 
}
\end{figure}

\begin{figure}
\centering
  \ForceWidth{4in}
\vspace*{-0.7cm}
  \hSlide{-1cm}
  \vSlide{-1cm}
  \BoxedEPSF{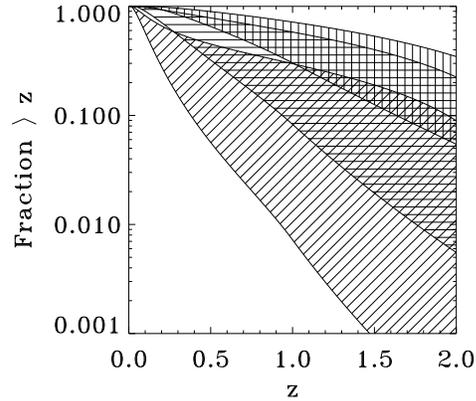}
  \vspace*{-1.7cm}
\caption{\label{fig:frac}
Predicted cumulative redshift distribution for surveys at
$100\kappa$ mJy (diagonal hatching), $10\kappa$ mJy (horizontal
hatching) and $1\kappa$ mJy (vertical hatching), where 
$\kappa=1\pm0.3$ indicates the uncertainty in the ELAIS PHOT absolute
flux calibration. The bounds of the shaded areas are dominated by the
assumed maximum redshift at which luminosity evolution occurs. 
}
\end{figure}

A much stronger 
constraint on the strength of the evolution comes from the
source counts slope. The $90\mu$m counts in the entire ELAIS areas
are significantly
non-Euclidean (Paper III), but the counts do show a suprising upturn
at the faintest end (figure \ref{fig:counts}). The cause of this
upturn is not 
clear, but it is too large to be entirely attributable to 
uncertainties in the completeness correction. One possibility is that
some of the faintest sources are spurious glitch events, but all the
sources have been eyeballed and accepted by at least two
observers. Such a glitch population would have to be somewhat
pathological. 
Evidence for comparably strong evolution at $<200$mJy in the Lockman
Hole is also reported by Matsuhara et al. (2000) based on model fits
to the fluctuations in their $90\mu$m maps, consistent with the
faintest ELAIS points. 
Nevertheless, in the absence of a clear model-independent
interpretation, 
we confine our source count fits to the $>200$mJy region. We note 
that the sources used in the luminosity function are all reliably
cross-identified at other wavelengths. The source counts at
$>200$mJy are in good agreement 
with the Lockman Hole counts from
Linden-V\o rnle et al. 2000 (figure \ref{fig:counts})
who use a very different source
extraction and flux calibration procedure to ELAIS. 
The 
ELAIS and Lockman 
counts for the faintest bin in figure \ref{fig:counts}
($165-294$mJy) are $27\pm4$ and
$22\pm9$ deg$^{-2}$ dex$^{-1}$ respectively, and in the brightest bin
($294-522$ mJy) 
$19\pm4$ and $4^{+9}_{-3.3}$ deg$^{-2}$ dex$^{-1}$ respectively. 
Note that 
this represents only a single source in the Lockman Hole, so
Poissonian errors would underestimate the error in this case. 
The $\pm1\sigma$
bounds in the case of $1$ observed source are $0.18-3.3$, and 
all the remaining errors quoted in the counts are
$\pm1\sigma$ Poissonian errors.

\begin{table}
\begin{tabular}{lll}
$\phi^* h_{65}^{-3}$ & $\log_{10}L^* h_{65}^{2}/L_\odot$ & $\alpha$\\
 &  & \\
$5.4\pm1.8$ & $9.67\pm1.47$ & $1.73\pm0.04$ \\
\end{tabular}
\caption{\label{tab:local_lf} Best-fit parameters of the local
luminosity function, with $\sigma=0.724$ assumed. The reduced $\chi^2$ 
of the best fit is $0.66$.}
\end{table}

If we {\it assume} the local luminosity function
derived above undergoes $(1+z)^\alpha$ luminosity evolution we
can obtain a constraint on the evolution parameter $\alpha$. Note that 
this is relatively insensitive to the flux calibration uncertainty
(Paper III). 
The correct relative normalisation between $90\mu$m and $100\mu$m for
these purposes is one which achieves continuity in the source
counts. The current $90\mu$m calibration satisfies this almost exactly 
(Paper III) in that the bright end of the $90\mu$m PHOT counts
dovetail with the faint end of the $100\mu$m IRAS counts, with little
room for error in the {\it relative} calibration. 
The Kolmogorov-Smirnov test on the shape of the source counts 
yields a
lower limit 
of $\alpha>1.5$ at $95\%$ confidence, and no-evolution models can
be excluded at the $98\%$ level. Stronger
constraints still are possible if we include the normalisation of the
source counts. Our goodness-of-fit statistic for this is $P_{\rm KS}
\times P_{\bar{\rm N}}$, where $P_{\rm KS}$ is the significance level of
rejecting the model of the shape of the source counts, and
$P_{\bar{\rm N}}$ is the significance level of rejecting the
predicted normalisation. The distribution function for such statistics 
is quoted in e.g. Dunlop \& Peacock 1990. We obtain
$\alpha=2.45\pm0.85$ ($95\%$ confidence). 
This is stronger than the constraint
obtained purely from a $V/V_{\rm max}$ analysis of the spectroscopic
sub-sample, partly because the number of galaxies comprising the
source counts is much larger, partly also because we have assumed
a fixed functional form for the zero redshift luminosity function, but 
mainly because we have set the $z=0$ normalisation from IRAS.

\section{Discussion and conclusions}\label{sec:discussion}
We have found that both the shape and normalisation of the
$90\mu$m luminosity function is well-described by the
local luminosity function of Rowan-Robinson, Helou and Walker (1987), 
with 
evolution consistent with $(1+z)^{2.45 \pm 0.85}$ pure luminosity
evolution, using 
the transformation $S_{100}\simeq S_{90}$. Pure density evolution is
already 
excluded in this population as it conflicts with sub-mJy radio source
counts (Rowan-Robinson et al. 1993). 

We can use these results to predict the redshift ranges of current and 
future $90\mu$m surveys. We assume the evolution extends to between
$z=1-2$ and is then unevolving. Figure \ref{fig:frac} shows the $95\%$
confidence limit for surveys of various depths, incorporating the
uncertainty in the ELAIS $90\mu$m absolute flux calibration. One
corollary is  that at least $10\%$ of the currently
unidentified $90\mu$m galaxies in ELAIS extend to $z>0.5$, possibly
even $z>2$. 

Is it reasonable to assume that star formation dominates the
bolometric luminosities of our $90\mu$m sample? 
Genzel et al. (1998), Lutz et al. (1998) and Rigopoulou et
al. (1999) have presented mid-infrared
spectroscopy of ultraluminous galaxies ($L\ge 10^{12} L_\odot$) in the
same redshift interval as our sample, 
and using the PAH features as a star formation indicator
found that star formation dominates the bolometric power outputs in
$70-80\%$ of ULIRGs. 
The AGN bolometric
fraction derived this way decreases monotonically at lower
luminosities (e.g. Lutz et al. 1998). 
In local IR galaxies between
$10^{11}$ and $10^{12} L_\odot$, $\sim15\%$ have Seyfert II spectra
(Telesco 1988), 
though care should be
taken in interpreting AGN dominance from optical spectra (e.g. 
Lutz et al. 1999, Taniguchi et al. 1999). 
At $L<10^{11}L_\odot$ most local IR galaxies are single, gas rich
spirals (e.g. Sanders \& Mirabel 1996) in which many lines of argument 
point to starburst dominance in the IR (e.g. Telesco 1988): for
example, the similarity of  
the spectral energy distributions to those of galactic star forming
regions; the linear $L_{\rm IR}-L_{\rm CO}$ correlation over many
orders of magnitude in luminosity; and optical/IR spectroscopic
confirmation of star formation activity. 
It therefore seems likely 
that our sub-ultraluminous population should not be powered 
by active nuclei, as is also suggested by the low fraction of
spectra
in our sample ($10\%$) with Seyfert II features.  
(This does not imply that AGN activity cannot be present at a 
weak or bolometrically-neglectable level, e.g. Ho, Filippenko \&
Sargent 1997.) 
At the lowest luminosities
there may be a significant contribution from cirrus which is at least
partly illuminated by the old stellar population in the galaxies
(Telesco 1988, Morel et al. 2000), though such galaxies nevertheless
also still obey the starburst radio-far-IR relation (e.g. Condon
1992). 
We conclude that the galaxies in our sample have their bolometric
luminosities dominated by star formation. 

The Saunders et al. (1990) local luminosity density implies a local
star formation rate of $0.023\pm0.002$ M$_\odot$ yr$^{-1}$ Mpc$^{-3}$ 
(Oliver et
al. 1998) assuming a
Salpeter initial mass function from $0.1$ M$_\odot$ to $125$
M$_\odot$. Our results imply this rises to  
$0.036\pm0.009$ M$_\odot$ yr$^{-1}$ Mpc$^{-3}$ by $z=0.2$ ($95\%$
confidence), significantly 
higher 
than the Tresse \& Maddox (1998) H$\alpha$-based estimate of
$0.022\pm0.007$ M$_\odot$ yr$^{-1}$ Mpc$^{-3}$. 
(Note that the two Seyfert II spectra in our sample 
are at the brightest end of the
luminosity function where the contribution to the luminosity density
is the smallest. The luminosity density is dominated by the
galaxies around or below the break in table \ref{tab:local_lf}.)
We believe 
this is due to two factors: firstly, the H$\alpha$-based estimates are 
not immune to extinction effects, even if performing Balmer decrement
reddening corrections (e.g. Serjeant, Gruppioni \& Oliver 1998);
secondly, the Tresse \& Maddox (1998) survey area is sufficiently
small to be affected by large-scale structure (e.g. Oliver, Gruppioni
\& Serjeant 1998). ELAIS is immune to these problems, as the cosmic
variance over the PHOT survey area is $<20\%$ at $z<0.3$ (Paper I). 
These star formation rates are however consistent with extrapolations
from those derived from mid-IR data at higher redshift by Flores et
al. (1999) and Rowan-Robinson (1997). 

The prospects for improving on the results presented here are
excellent. A large-scale spectroscopic follow-up of the ELAIS northern
areas is currently underway, which will have a major impact on the
science analysis of ELAIS. In particular, the ELAIS $15\mu$m 
luminosity function will probe higher redshifts than accessible to
PHOT at $90\mu$m, from which the $15\mu$m luminosity density can 
constrain the cosmic star formation history (e.g. Serjeant 2000). 


\section*{Acknowledgements}
We would like to thank Dave Clements and the anonymous referee 
for helpful comments on this
paper. 
This work was
supported by PPARC (grant 
number GR/K98728) and by the EC TMR Network programme
(FMRX-CT96-0068).

\section*{Appendix: $1/V_{\rm max}$ with multivariate flux limits}
In a volume-limited non-evolving sample, the comoving number density of   
objects can be calculated trivially:
\begin{equation}
\label{eqn:1}
\phi = \frac{N}{V} = \frac{1}{V}\sum_{i=1}^{N} 1
\end{equation}
where $V$ is the volume and $N$ the number of objects. Suppose that
the sample is incomplete in whatever way, and that the probability of
the $i^{\rm th}$ object being contained in the sample is
$p_i$. Provided that none of the $p_i$s in the underlying galaxy
population are zero, the comoving number density in an observed sample 
of $N_{\rm obs}$ objects can be expressed as
\begin{equation}
\label{eqn:2}
\phi = \frac{1}{V}\sum_{i=1}^{N_{\rm obs}} p_{i}^{-1} .
\end{equation}
This will be an unbiased estimator of the underlying value. 

Applying this formalism to our current data set, the incompleteness is 
due to the multivariate flux limits. We treat the (non-evolving)
galaxies as sampling random redshifts within the volume, so that each
$p_i$ is the probability that such a galaxy would lie above the flux
limits. The statement that the underlying $p_i$s are all non-zero is
equivalent to assuming that there are no galaxies (in the $90\mu$m
luminosity range being considered) that would lie outside the
selection criteria 
at all redshifts. We can be confident that this is the case, because
any sufficiently local galaxy will have an optical identification
passing criterion (a) of our selection function. A hypothetical
population of FIR-luminous, optically-faint galaxies can already be
excluded from IRAS. 

There are two equivalent ways of calculating the $p_i$s. If we embed
our flux-limited sample in a larger volume of size $V_0$, we can use
the assumption that the population is non-evolving and the fact that
the flux limits are monotonic in $z$ to express the $p_i$s as 
\begin{equation}
\label{eqn:3}
p_i = \frac{V_{{\rm max}, i}}{V_0}
\end{equation}
where $V_{{\rm max}, i}$ is the volume enclosed by the maximum
redshift $z_{{\rm max}, i}$ at which the $i^{\rm th}$ object is
visible. Note that this is the smallest redshift at which the object
fails {\it any} of the selection criteria. The number density in a
sample of $N_{\rm obs}$ galaxies is then
\begin{equation}
\label{eqn:4}
\phi = \frac{1}{V_0} \sum_{i=1}^{N_{\rm obs}} \frac{V_0}{V_{{\rm max}, i}}
     = \sum_{i=1}^{N_{\rm obs}} \frac{1}{V_{{\rm max}, i}} .
\end{equation}
The RMS error is simply
\begin{equation}
\label{eqn:5}
\Delta\phi = \sqrt{\sum_{i=1}^{N_{\rm obs}} \frac{1}{V^2_{{\rm max}, i}}} .
\end{equation}
An alternative but equivalent method of calculating the $p_i$s is to
model the incompleteness due to some or all of the multivariate flux
limits by introducing a weighting factor to the differential volume
elements:
\begin{equation}
\label{eqn:6}
dV' = \gamma(z, ...) dV .
\end{equation}
The $z_{{\rm max}, i}$ values would then be calculated using the
remaining un-modelled flux limits only. If all the flux limits have
been modelled, then $\forall i ~ z_{{\rm max}, i}=z_0$ where
$V(z_0)=V_0$. The $\gamma$ factor would drop to zero at some redshift
if (under the previous method for calculating the $z_{{\rm max},i}$)
$z_0>\max(z_{{\rm max}, i})$.  

Here, our approach is to use the first method (equation \ref{eqn:4})
to treat the multi-variate flux limits, as it is less
model-dependent. However we use a weighting function on the volume
elements to correct for the ELAIS $90\mu$m completeness function
(figure 6 of Paper III):
\begin{equation}
\label{eqn:7}
V'_{{\rm max}, i} = \int_{0}^{z_{{\rm max}, i}} 
			\gamma(z, S_{i}) \frac{dV}{dz} dz
\end{equation}
where $S_i$ is the $90\mu$m flux of the $i^{\rm th}$ object. 

To calculate the luminosity function in a redshift bin, an additional
top-hat selection function is applied in redshift space and
incorporated into the calculation of the $z_{{\rm max}, i}$. The
volume enclosed by the minimum redshift of the bin must also be
subtracted from the $V_{{\rm max}, i}$ in equation \ref{eqn:3}. We are
currently confident that none of the underlying $p_i$s are zero, but
if we calculated the luminosity function in a series of redshift bins 
this would not necessarily be the case for all bins. For example,
there are galaxies in table \ref{tab:sample} which only pass criterion 
(a) at the lowest redshifts, and these may be unobservable in
principle in higher redshift bins. We would then need to make some
model of the multi-variate correlations to correct for the missing
$p_i$=0 galaxies. We therefore restrict ourselves to a single redshift 
range (de-evolving the galaxies if necessary), partly to avoid the
model-dependence of correcting for missing populations, but partly
also because of our small sample size.

\end{document}